\documentclass[12pt,preprint]{aastex}

\newcommand{\fig}[1]{Figure~\ref{#1}}
\newcommand{\tbl}[1]{Table~\ref{#1}}
\newcommand{\speed}[1]{#1 km s${}^{-1}$}
\newcommand{\aspeed}[1]{$\sim$#1 km s${}^{-1}$}
\newcommand{\rsun}[1]{${#1}\,R_\sun$}

\shorttitle{}
\shortauthors{Shen et al.}
\begin{document}

\title{OBSERVATIONAL STUDY OF THE QUASI-PERIODIC FAST PROPAGATING MAGNETOSONIC WAVES AND THE ASSOCIATED FLARE ON 2011 MAY 30}
\author{Yuandeng Shen\altaffilmark{1,2} and Yu Liu\altaffilmark{1}}
\altaffiltext{1}{Yunnan Astronomical Observatory, Chinese Academy of Sciences, P.O. Box 110, Kunming 650011, China; ydshen@ynao.ac.cn}
\altaffiltext{2}{Graduate University of Chinese Academy of Sciences, Beijing 100049, China}

\begin{abstract}
On 2011 May 30, quasi-periodic fast propagating (QFP) magnetosonic waves accompanied by a C2.8 flare were directly imaged by the Atomospheric Imaging Assembly instrument on board the {\em Solar Dynamics Observatory}. The QFP waves successively emanated from the flare kernel, they propagated along a cluster of open coronal loops with a phase speed of \aspeed{834} during the flare's rising phase, and the multiple arc-shaped wave trains can be fitted with a series of concentric circles. We generate the $k$--$\omega$ diagram of the Fourier power and find a straight ridge that represents the dispersion relation of the waves. Along the ridge, we find a lot of prominent nodes which represent the available frequencies of the QFP waves. On the other hand, the frequencies of the flare are also obtained by analyzing the flare light curves using the wavelet technique. The results indicate that almost all the main frequencies of the flare are consistent with those of the QFP waves. This suggests that the flare and the QFP waves were possibly excited by a common physical origin. On the other hand, a few low frequencies revealed by the $k$--$\omega$ diagram can not be found in the accompanying flare. We propose that these low frequencies were possibly due to the leakage of the pressure-driven $p$--mode oscillations from the photosphere into the low corona, which should be a noticeable mechanism for driving the QFP waves observed in the corona.
\end{abstract}


\section{INTRODUCTION}
The magnetically dominated plasma of the solar corona, which is an elastic and compressible medium, can support the propagation of various magnetohydrodynamic (MHD) waves (e.g., slow mode, Alfv$\rm \acute{e}$n, and fast mode waves). In the last decade, observations from ground-based and space-borne instruments have led to the detection of various waves in the solar atmosphere \citep[see][]{demo02b,naka05,bane07,demo09}. For example, (1) oscillations or standing waves of slow and fast modes in a variety of coronal magnetic structures \citep{asc99,schr99,wang02,naka03,okam07,ofma08}, (2) propagating compressible slow magnetosonic waves in solar polar plumes \citep{defo98,ofma97,ofma99} and in coronal loops \citep{berg99,demo00,demo02a,mars03,robb01,wang09a,wang09b}, (3) transverse incompressible Alfv$\rm \acute{e}$n waves in the corona \citep{cirt07,depo07,tomc07,tomc09,jess09}, and (4) compressible fast magnetosonic waves in coronal loops \citep{will01,will02,mcla09,liu11,ofma11}. It should be noted that the Alfv$\rm \acute{e}$n wave interpretation of the observations in \cite{tomc07} could alternatively be explained as kink waves \citep{vand08}. The single-pulse ``EIT waves'' \citep{thom98} with typical speeds of 200--\speed{400} were often interpreted as fast magnetosonic waves \citep{wu01,ofma02,vero10,liu10} and thought to be the coronal counterparts of the chromospheric Moreton waves \citep{more60}, which have been successfully explained as coronal shock waves sweeping the chromosphere \citep{uchi68}. However, the wave interpretations for EIT waves have not yet been fully accepted by many other authors for some drawbacks in the wave models \citep[see][]{dela99,dela08,klas00,attr07,attr09,will06,warm10,chen02,chen05,chen11}. Very recently, the universal wave interpretations for coronal oscillations are challenged by the discovery of the rapid quasi-periodic mass up-flows at transition-region and coronal temperatures \citep{saka07,depo09,depo10,mcin09a,mcin09b,mcin10,mcin11,tian11}, and those up-flows were suggested to provide hot plasma into the corona and thereby heating the corona or as a possible source for the fast solar wind \citep{depo09,depo11,mcin09b,pete10,depo10,hans10,tian11}.

On the other hand, the oscillations and quasi-periodic pulsations (QPPs) of solar flares have also been detected for many years in wide wavelength bands ranging from radio to gamma-ray, on timescales from fractions of seconds to several minutes \citep[e.g.,][]{park69,asch87,flei02,foul05,kisl06,tan08,naka09,naka10,kupr10,tan2012}. QPPs can usually be observed in light curves of solar flares in various wavelength bands during the entire flaring process. Moreover, they can also be observed in stellar flares with a period of several minutes \citep{math03,math06}. In different wavelength bands, QPPs often shows a synchronous evolvement patten in phase \citep{asai01,ingl08}. In some events, several significant periods could be detected simultaneously in one single flare \citep{kisl06}. For different timescales of QPPs in solar flares, \cite{naka09} split them into short (sub-second) and long period QPPs. The short period QPPs are usually observed in radio emission, and are likely associated with the interaction of electro-magnetic, plasma or whistler waves with accelerated particles \citep[e.g.,][]{asch87}, while the long period QPPs are usually detected in microwave, white light, EUV and X-ray emissions, and are possibly associated with MHD processes \citep{naka09}. \cite{naka04} performed a numerical study on the evolution of a coronal loop in response to an impulsive energy release, finding that the QPPs of the loop density is associated with the second harmonic of a standing slow magnetosonic wave. In another publication \citep{tsik04}, it is further pointed out that the generation of the QPPs of the loop density is independent of the location of the heat deposition in the loop. Although there are many observational and theoretical studies in the past several decades, the underlying physical mechanisms that generate QPPs in solar flares still remain an open question. Up to the present, there are several possible interpretations for this issue, including modulation of electron dynamics by MHD oscillations \citep{zait82,flei08}, periodic triggering of energy releases by MHD waves \citep{robe83,robe84,foul05,naka06,ingl08,sych09}, MHD flow overstabilities \citep{ofma06}, and oscillatory regimes of magnetic reconnections \citep{klie00,mcla12}.

Quasi-periodic propagating fast magnetosonic waves with high speed remain the least observed among all coronal MHD waves. The scarcity of fast wave observations was mainly due to the instrumental limitation in the past. For wavelengths that are comparable with the characteristic size of coronal magnetic structures (e.g., length of coronal loops, width of coronal polar plumes), the typical periods of the waves are in the range from a few seconds to several minutes \citep{naka05}. Hence both spatial and temporal resolutions are necessary ingredients for the detection of fast magnetosonic waves. Namely, the time cadence must be shorter than the periods and the pixel size must be smaller than the wavelength of the fast magnetosonic waves. The first imaging observation that had been interpreted in terms of propagating fast magnetosonic wave had been reported by \cite{will01,will02}. They successfully imaged a propagating fast magnetosonic wave during the total solar eclipse on 1999 August 11, in which the wave travels through the apex of an active region coronal loop with a speed of \speed{2100}. \cite{coop03} and \cite{naka03} modeled the evolution of the propagating fast magnetosonic wave and confirmed the formation of the quasi-periodic wave trains predicted by \cite{robe83,robe84}. Very recently, \cite{liu11} reported the quasi-periodic fast propagating (QFP) magnetosonic waves in the low corona using the EUV observations taken by the Atmospheric Imaging Assembly instrument \citep[AIA;][]{leme12} on board the {\em Solar Dynamics Observatory} ({\em SDO}). The authors found that the multiple arc-shaped wave trains emanated consecutively near the associated flare kernel and propagated outward along a funnel of coronal loops with a phase speed of \aspeed{2200}. In addition, they also found that the main period of the associated QPPs flare was almost the same with the strongest period (181 s (5.5 mHz)) of the QFP waves. They thus concluded that the flare and the QFP waves have a common origin of possibly quasi-periodic magnetic reconnection. A three-dimensional numerical simulation study of the QFP waves reported by \cite{liu11} has been presented by \cite{ofma11}, in which they successfully reproduced the multiple arc-shaped wave trains that have similar amplitude, wavelength, and propagation speeds as reported by \cite{liu11}. They also discussed the possible excitation mechanism of the QFP waves and the applications of the observations for coronal seismology. Other numerical simulations of fast waves can be found in \cite{bog03}, \cite{hegg09}, \cite{fedu11}.

In this paper, we present an observational study of a QFP wave event that occurred on 2011 May 30. It was accompanied by a {\em GOES} C2.8 flare in NOAA active region AR11227 (S19E57) and a faint coronal mass ejection (CME) observed in white light. We find that the QFP wave trains consecutively emanated from the flare kernel and propagated along a bunch of open active region coronal loops which extended from AR11227 to the heliosphere, while another propagating wave are observed in the nearby closed transequatorial loops that connect the neighbouring active region AR11226 and the outskirts of AR11228 in the northern hemisphere. Using high spatial and temporal observations taken by AIA, we first generate the $k$--$\omega$ diagram of the QFP waves with Fourier transform, and then isolate the periods of the associated pulsation flare with light curves in various wavelength bands. By comparing the periods (frequencies) of the wave and the flare, we try to obtain some new clues about the driving mechanism of the QFP magnetosonic waves. Observations are described in Section 2, results are presented in Section 3, discussions and conclusions are given in Section 4.

\section{OBSERVATIONS}
The AIA on board {\em SDO} is propitious to detect fast propagating features such as fast magnetosonic waves of short periods. The telescope captures image of the Sun's atmosphere out to \rsun{1.3} and has high time resolution of up to 12 s. AIA has high spatial resolution of $1\arcsec.2$ in seven EUV and three UV--visible channels, which cover a wide temperature range from $\log T=3.7$ to $\log T=7.3$. All these parameters of AIA are necessary ingredients for detecting fast propagating waves in the low corona. In the event analyzed here, the QFP waves along open active region coronal loops can only be observed in the AIA 171 \AA\ (\ion{Fe}{9}; $\log T=5.8$) observations, which mainly image the quiet corona and the upper transition region. In the meantime, another wave that was trapped in closed transequatorial loops can be detected at both 171 \AA\ and 193 \AA\ (\ion{Fe}{12}; $\log T=6.1$) lines. The {\em Reuven Ramaty High Energy Solar Spectroscopic Imager} \citep[{\em RHESSI};][]{lin02} hard X-ray (HXR) count rates in the energy bands (4 s integration) of 12--25 keV and the {\em Geostationary Operational Environmental Satellite} ({\em GOES}) soft X-ray (SXR) flux are also used to analyze the periods of the accompanying QPPs flare. All images used in this paper are differentially rotated to a reference time (11:00:00 UT), and solar north is up, west to the right.

\section{RESULTS}
The event presented in this paper includes the QFP waves along the open coronal loops that rooted in AR11227, and another fast magnetosonic waves in the closed transequatorial loops that connect the neighbouring active region AR11226 (S21E42) and the outskirts of AR11228 (N17E55) in the northern hemisphere. These waves were accompanied by a {\em GOES} C2.8 flare that occurred in AR11227 on 2011 May 30. The start, peak, and end times of the flare are 10:48, 10:57, and 10:59 UT respectively. According to the {\em RHESSI} flare record, the heliographic coordinates of the flare kernel were $x = -787\arcsec, y = -283\arcsec$. In addition, a faint CME in white light was also observed to be associated with a jet-like plasma eruption which occurred in AR11227 just before the flare and the waves.

\begin{figure}
\epsscale{0.7}
\plotone{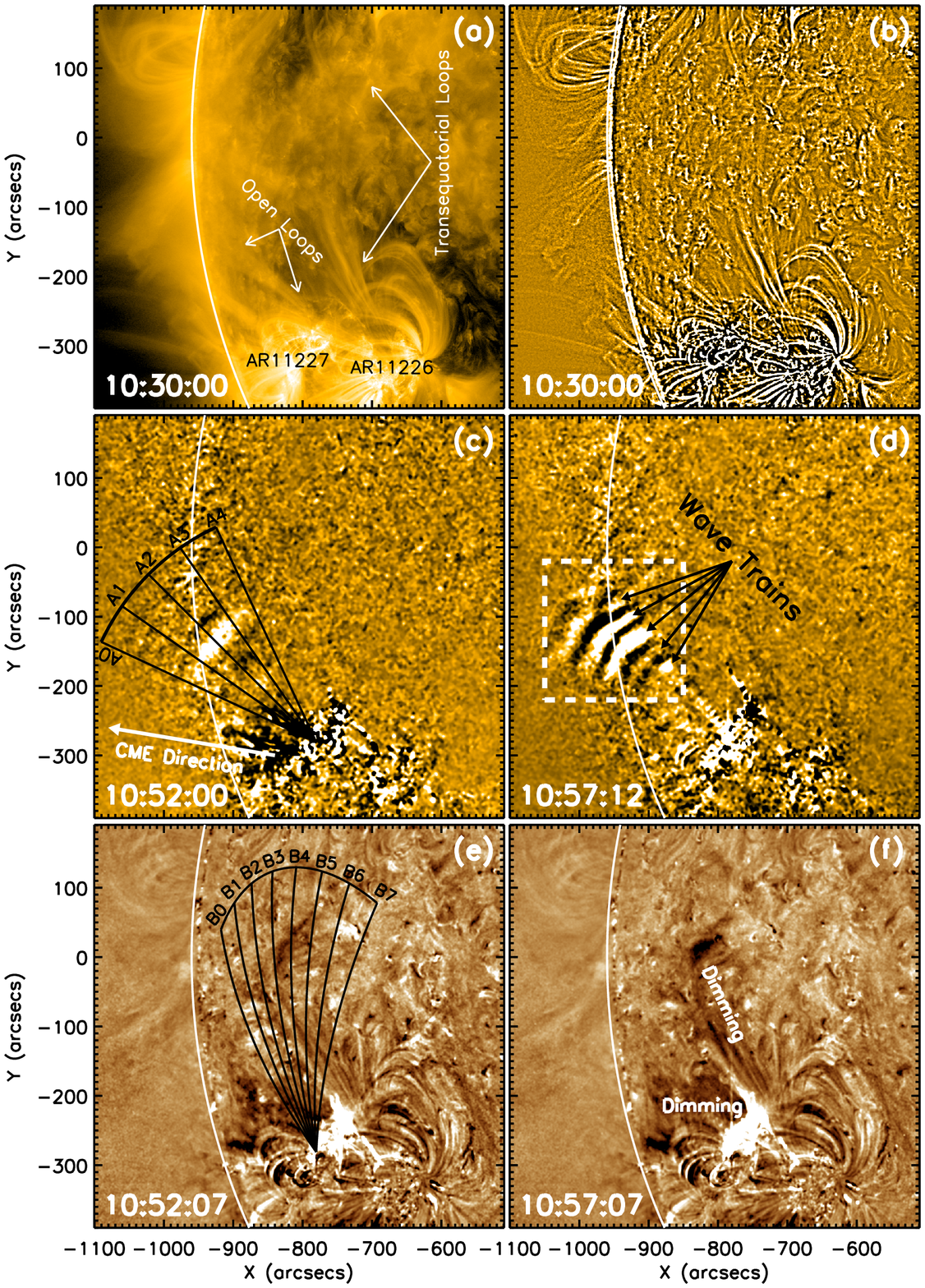}
\caption{An overview of the wave event. (a) AIA 171 \AA\, (b) the filtered image of (a), which is obtained by subtracting a smoothed image with a boxcar average over 15 $\times$ 15 pixels, (c)--(d) AIA 171 \AA\ running difference images (Animation 1), and (e)--(f) AIA 193 \AA\ base difference images (Animation 2). Arrows in frame (a) point to the open and closed transequatorial loops which guide the propagating waves, while jointed black arrow in frame (e) indicate the multiple arc-shaped wave trains. The white long arrow in frame (c) indicates the direction of the associated CME. The white dashed box in frame (e) indicates the selected region where Fourier analysis is applied. The coronal dimming regions behind the waves are also indicated in frame (f). Sectors A0--A4 in frame (c) and B0--B7 are used to obtain the time-distance diagrams along the propagating directions of the propagating waves, and the time-distance diagrams are shown in \fig{f2} and 7, respectively. The white curve in each frame marks the limb of the solar disk (the same in the following figures). The FOV is $590\arcsec \times 580\arcsec$ for each frame. \label{f1}}
\end{figure}

An overview of the event is displayed in \fig{f1}. One can see that a cluster of funnel-like open loops rooted in AR11227 and extended northeast off the disk limb, and another bunch of closed transequatorial loops connected AR11226 and the outskirts of AR11228 in the northern hemisphere. These loops can be identified clearly on the raw and the filtered AIA 171 \AA\ images (see \fig{f1}(a) and (b)). The QFP waves were observed along the open loops. They can be distinguished at about 10:50:00 UT and then multiple arc-shaped wave trains are noticed from the 171 \AA\ running difference images (see \fig{f1}(c) and (d), and Animation 1 available in the online version of the journal). The wave trains consecutively and alternately emanated from the flare kernel and exhibited an arc-like shape. They propagated northeastward along the open loops and finally dimmed off the disk limb successively. This suggests that the QFP waves were not propagating surface waves as the Moreton and EIT waves, in agreement with the results by \cite{liu11}. The start time of the QFP waves (10:50 UT) was slightly lagging the beginning of the accompanying flare (10:48 UT), and the end time of the QFP waves (11:02 UT) was also lagging the flare peak time (10:57 UT) several minutes. This close temporal relationship between the QFP waves and the flare implies that they might be closely connected. In addition, we note that the QFP waves could only be observed in AIA 171 \AA\ observations, and the open loops that guided the QFP waves could not be observed in the other wavelength bands. These results indicate the subtle temperature dependence of the QFP waves.

It should be noted that a jet-like plasma ejection around the base of the open loops was observed before the beginning of the start of the accompanying flare and the QFP waves. This small plasma ejection started at about 10:30 UT and ended at about 10:53 UT. It erupted outward eastwardly and it possibly resulted in the faint CME due to the intimate temporal and spatial relationship between them (see the long white arrow in \fig{f1}). Furthermore, the obvious coronal dimming region close to the source region of the plasma ejection was possibly the on-disk signature of the occurrence of the CME. The discrepancies on the start times, the propagating directions of the QFP waves and the jet-like plasma eruption together indicate that the two phenomena did not have a cause and effect relationship. In this paper, we mainly focus on the relationship between the QFP waves and the associated flare.

\begin{figure}
\epsscale{0.90}
\plotone{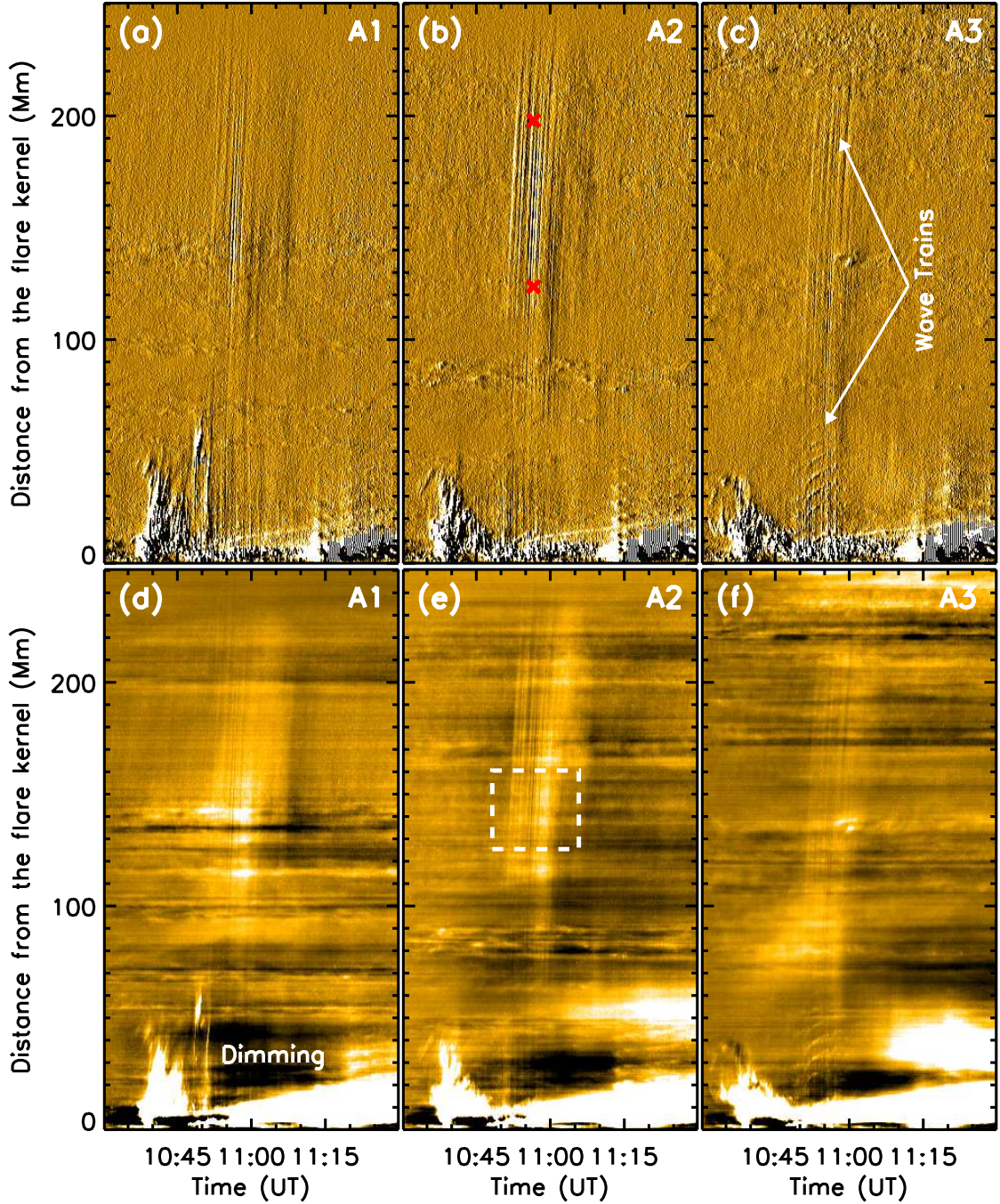}
\caption{Time-distance diagrams and analysis of the QFP waves along the open coronal loops. (a)--(c) Running difference time-distance diagrams obtained from AIA 171 \AA\ running difference images along the sectors A1--A3 shown in \fig{f1}(c), (d)--(f) the same as (a)--(c), but they are obtained from the AIA 171 \AA\ base difference images. The white arrows indicate the QFP wave trains in the time-distance diagram.  \label{f2}}
\end{figure}

\begin{figure}
\epsscale{0.80}
\plotone{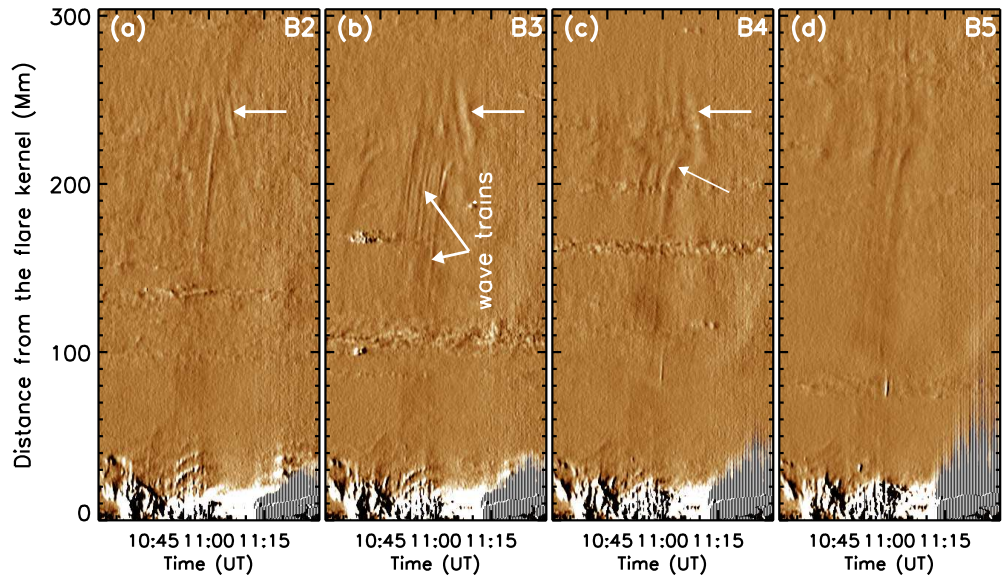}
\caption{Time-distance diagrams for analyzing the waves guided by the closed transequatorial loops. (a)--(d) AIA 193 \AA\ running difference time-distance diagrams obtained from the sectors B2--B5 shown in \fig{f1}(e). The white horizontal and the oblique arrows point to the reflected wave trains, while the jointed white arrows in frame (b) indicate the multiple propagating wave trains. \label{f7}}
\end{figure}

For the multiple arc-shaped wave trains propagating along the open loops, each of them can be fitted with a circle whose center is located at the site of the flare kernel. To analyze the kinematics of the QFP waves, we place five lines (A0--A4) in a sector region where the QFP waves were propagating. The angle between the adjacent lines is $10\degr$ (see \fig{f1}(c)). By averaging 10 pixels across each line and then stacking the obtained profiles in time sequence, the time-distance diagrams along the line can thus be obtained, and the results are shown in \fig{f2}. For the wave guided by the closed loops, it propagated resembling a surface wave that was confined on the solar surface(see Animation 2). In addition, obvious coronal dimming region can be identified following the wave fronts (see \fig{f1}(e) and (f)). To minimize human subjectivity and the spherical projection effect, we select the flare kernel as the new ``north pole'' and the angle between the new heliographic ``longitudes'' is still $10\degr$ (see \fig{f1}(e)). Along each ``longitudinal'' great circle, we obtain the intensity profile of each image by averaging 10 pixels in the ``latitudinal'' direction. By stacking these intensity profiles over time, one can obtain a two-dimensional time-distance diagram, as shown in \fig{f7}.

\fig{f2} shows the time-distance diagrams obtained from AIA 171 \AA\ running (\fig{f2}(a)--(c)) and base (\fig{f2}(d)--(f)) difference images along cuts A1--A3 shown in \fig{f1}(c). These time-distance diagrams exhibit the kinematics of the QFP waves guided by the open coronal loops. From the running difference time-distance diagrams, one can see that the multiple wave trains are shown as a series of steep stripes with positive slope representing the propagating speed of the QFP waves. It can be seen that these steep stripes are originated from the flare kernel and become prominent at a distance of about 100 Mm far from the flare kernel. It should be noted that the QFP waves underwent a deceleration phase before they became prominent. Since the solar corona is a highly inhomogeneous medium in nature \citep{ofma11}, the deceleration of the QFP waves was possibly resulted from the changing of the properties of the propagation medium (such as the variation of density and magnetic field strength along with height of the loops). The speed of each stripe in the plane of the sky is measured using a linear fit method. Within the distance ranging from 120 to 220 Mm far from the flare kernel, within which the wave trains were mostly prominent, we find that the speeds of the different stripes show little difference along different lines, and the average speed of all wave trains is about \aspeed{774}. For the wave trains within the distance less than 100 Mm far from the flare kernel, the measurement is inaccurate since the wave trains are very obscure. For a rough estimation, the speed of the wave trains within such a distance should be higher than \speed{1000}. It should be noted that a dimming region is detected on the base difference time-distance diagrams close to the flare kernel (see the bottom row in \fig{f2}). As mentioned above, it was associated with the jet-like plasma ejection and the CME observed in white light, and it had probably no direct relation with the QFP waves studied here.

\begin{figure}
\epsscale{0.80}
\plotone{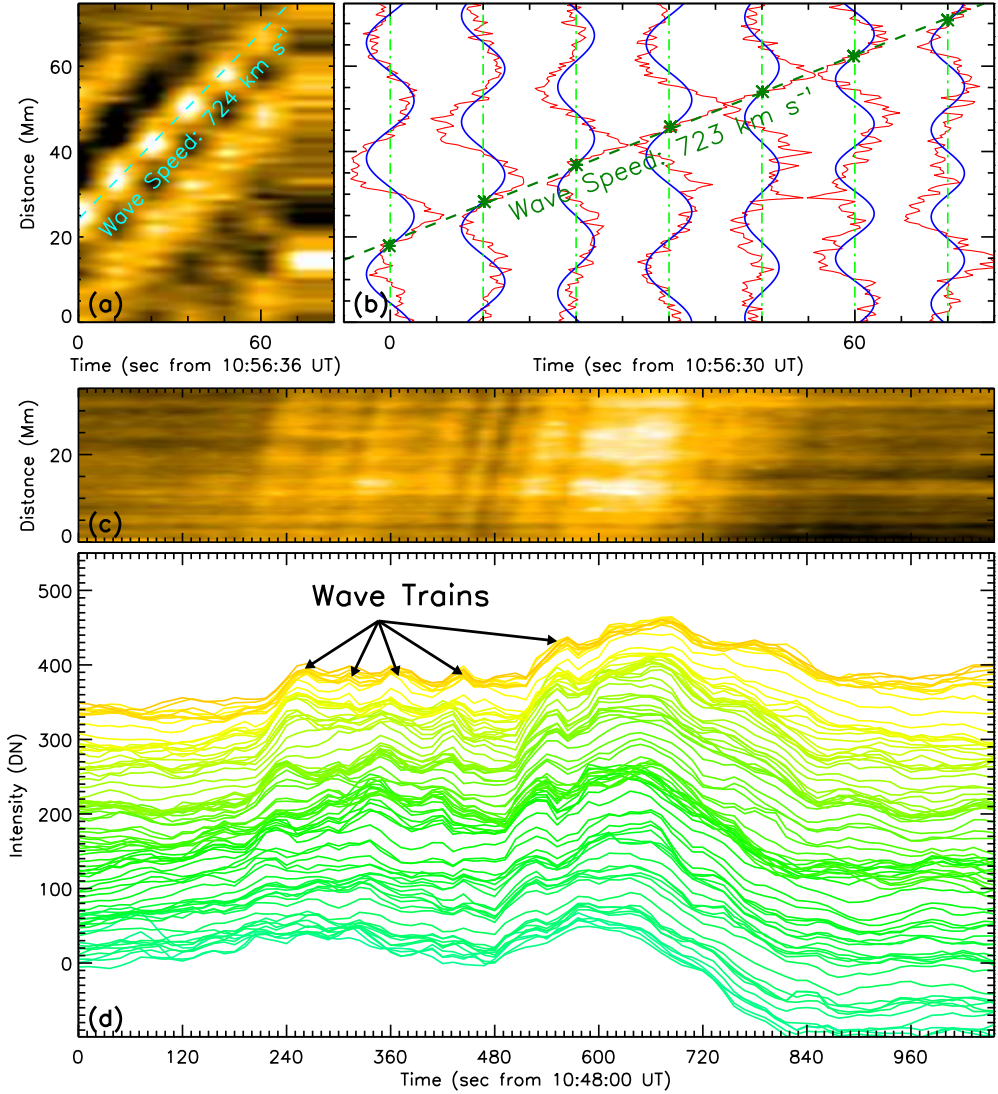}
\caption{(a) Subregion of \fig{f2}(b) at times and distances marked by the two red ``$\times$'' signs. The horizontal direction is magnified by 12 times. (b) the intensity profile of each pixel ($x$-axis) as a function of distance ($y$-axis) at seven consecutive times from 10:56:30 UT (red curves). Each profile is normalized into [-6,6] DN and then it is incrementally shifted by 12 DN along the $x$-axis, which equals to the time cadence of the AIA 171 \AA\ filter and thereby the $x$-axis also serves as time axis. The blue curves are the fitted curves of a sine function. (c) the dashed box region shown in \fig{f2}(e), which is also magnified in the $x$-axis direction to fit the panel. (d) the intensity profile of each pixel ($y$-axis) as a function of time ($x$-axis). Note that the intensity profile of larger distance is stacked on the previous one. Each curve is shifted with 5 DN in $y$-axis direction. \label{f3}}
\end{figure}

We further check the intensity variation of the QFP wave trains along and cross the lines that are used to obtain the time-distance diagrams, and the results are shown in \fig{f3}. We cut a narrow subregion from the running difference time-distance diagram along A2 (\fig{f3}(a)). The time interval for this subregion is from 10:56:36 UT to 10:58:00 UT, corresponding to seven time points of the AIA's 12 s time cadence. On this sub time-distance diagram, the wave trains are more clear. By making a linear fit to the wave trains, we obtain the speed of the propagating wave which is \aspeed{724}. The intensity profile at each time point as a function of distance shows obvious quasi-periodic variation as indicated by the red curves in \fig{f3}(b), in which each profile is normalized into [-6,6] DN and then incrementally shifting by 12 DN that corresponds to the time cadence of the AIA instruments. Thus the $x$--axis also servers as the elapsed time. It is interesting that each intensity profile can be fitted with a sine function $Asin[2\pi(r-r_0)/\lambda]$ (blue curve), here $A$ being the amplitude, $\lambda$ the wavelength, and $r_0$ the initial phase in distance. The average fitted parameters are $\lambda=23.8$ Mm, $A=2.58$ DN. Based on these fitted curves, we can obtain the phase speed of the QFP waves in the plane of the sky, which is \aspeed{723}, almost the same with the phase speed measured from the sub time-distance diagram shown in \fig{f3}(a). To investigate the intensity variation across the lines, a subregion is cut from the base difference 171 \AA\ time-distance diagram obtained from A2 (dashed box shown in \fig{f2}(e)), and the intensity profiles at each position is plotted as a function of time, in which the next profile of larger distance is stacked on the previous one. The QFP wave trains are more clear, and obvious periodic variation of intensity can be found (see the black arrows in \fig{f3}(d)).

\begin{figure}
\epsscale{0.80}
\plotone{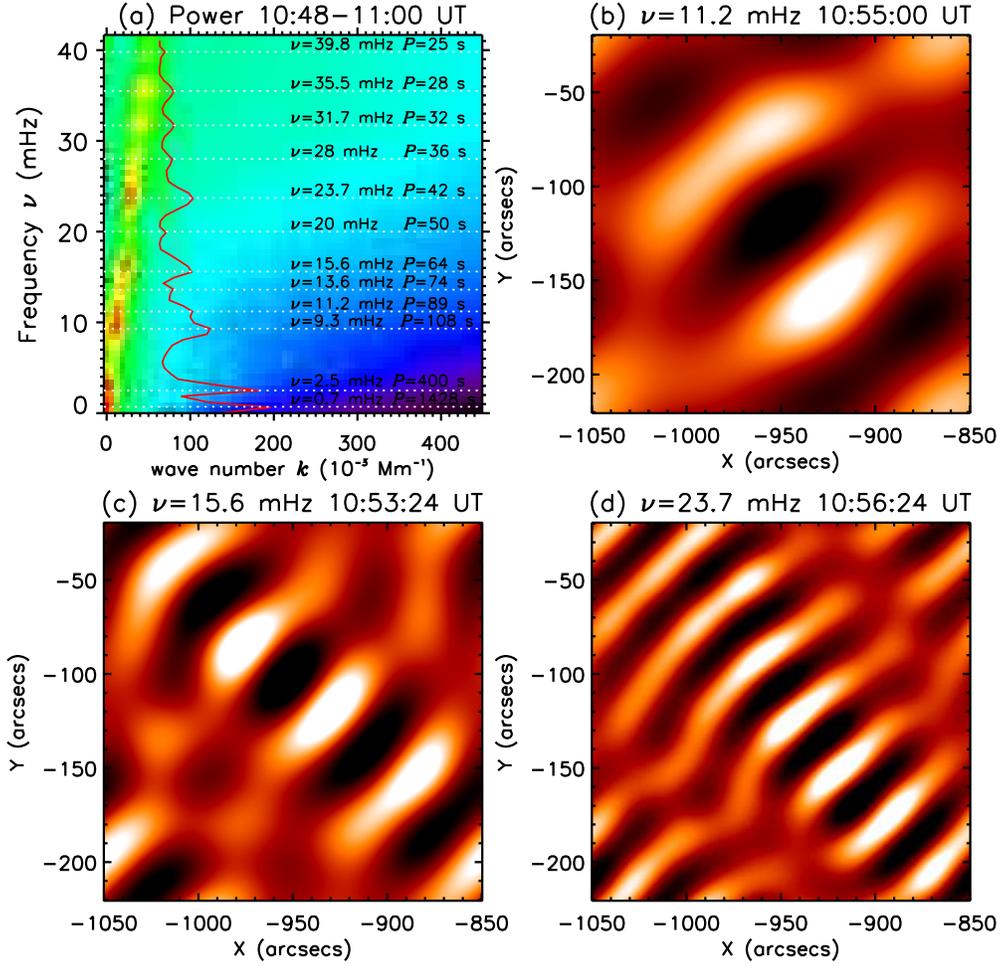}
\caption{Fourier analysis of the QFP waves guided by the open loops. (a) Fourier power ($k$--$\omega$ diagram) of a three-dimensional data tube of 171 \AA\ running difference images during 10:48--11:00 UT in the dashed box region shown in \fig{f1}(d). The red curve is the normalized intensity profile of the straight ridge. The white dotted lines indicate the possible frequencies of the QFP waves, which are determined by the peaks showing on the intensity profile of the straight ridge. (b)--(d) AIA 171 \AA\ running difference sample images in the dashed box region shown in \fig{f1}(d), they are Fourier filtered with a narrow Gaussian function centered at the peak frequencies of 11.2 mHz, 15.6 mHz, and 23.7 mHz respectively. The FOV for frames (b)--(d) is $200\arcsec \times 200\arcsec$. \label{f4}}
\end{figure}

For MHD waves generated impulsively, we often observe the propagating wave packets in imaging observations. These wave packets represent a Fourier integral over all frequencies. Therefore, one can decompose these frequencies with the method of Fourier transform. In line with this thought, we first extract a three-dimensional data cube in the time domain (i.e., in ($x,y,t$) coordinates, the time interval is from 10:48--11:00 UT, corresponding to the rising phase of the associated flare), then transform the data cube into the frequency domain (i.e., in ($k_{\rm x},k_{\rm y},\nu$) coordinates, where $k_{\rm x}$ and $k_{\rm y}$ are wavenumbers along $x$ and $y$, $\nu$ is frequency). Then, the Fourier power is summed in the azimuthal $\theta$ direction of cylindrical coordinates ($k,\theta,\nu$), where $k=\sqrt{k_{\rm x}^2+k_{\rm y}^2}$. This yields a $k$--$\omega$ diagram of wave power as shown in \fig{f4}(a) \citep[also see][]{defo04,liu11}. The resolution of the $k$--$\omega$ diagram is $\Delta k= 6.83\times10^{-3}$ Mm$^{-1}$, $\Delta \nu=1.39$ mHz, which are calculated based on the FOV of the selected region and the time interval of the data cube. An obvious liner steep ridge that represents the dispersion relation of the QFP waves can be found in the $k$--$\omega$ diagram (see \fig{f4}(a)). This ridge can be well fitted with a straight line passing through the origin up to the Nyquist frequency of 41.7 mHz. This gives the phase ($v_{\rm ph}=\nu/k$) and the group ($v_{\rm gr}=d\nu/dk$) velocities of the QFP waves (\aspeed{834}). It should be noted that many prominent nodes can be found on this bright ridge, and they represent the available frequencies (periods) of the QFP waves. We also plot the normalized intensity profile of this bright ridge in \fig{f4}(a) as a red curve, from which we find the intensity peaks that represent the available frequencies (periods) of the QFP waves (see the horizontal dotted lines in \fig{f4}(a)). We find that the frequencies (periods) of the QFP waves range from 0.7--39.8 mHz (1428--25 s). \fig{f4}(b)--(d) are the snapshots of the Fourier filtered AIA 171 \AA\ running difference images with a narrow Gaussian function centered at the peak frequencies of 11.2, 15.6, and 23.7 mHz, which highlight the corresponding QFP wave trains at different frequencies (also see Animation 3 available in the online version of the journal). Obviously, larger frequency corresponds smaller distance between adjacent wave trains, namely shorter wavelength $\lambda$, since the propagating speeds of them are the same.

\begin{figure}
\epsscale{0.80}
\plotone{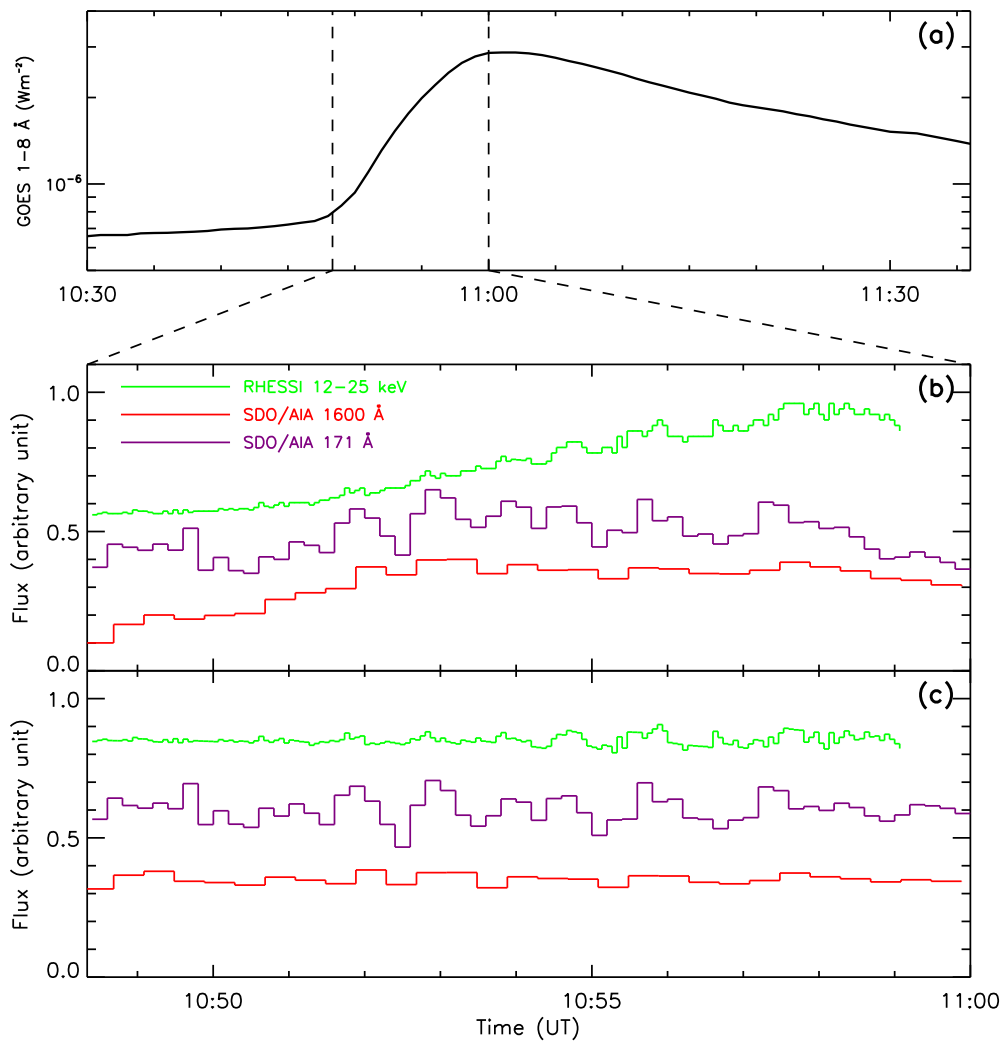}
\caption{Sample light curves of the accompanying flare. (a) time profile of {\sl GOES} 1--8 \AA\ SXR. (b) {\em RHESSI} HXR count rates in the energy band (4 seconds integration) of 12--25 keV (green), and AIA 171 \AA\ (purple) and 1600 \AA\ (red) light curves measured from the flare region. The time range is from 10:48--11:00 UT. (c) the same with (b), but for detrended fluxes obtained by subtracting the smoothed fluxes using a 120 s boxcar.  \label{f5}}
\end{figure}

Previous studies have indicated that an obvious source of propagating fast magnetosonic wave can be a flare \citep{robe84}. So we measure the light curves of the associated flare from the AIA observations by summing the intensities within a small region centered on the flare kernel. We show the sample light curves of 1600 \AA\ and 171 \AA\ in \fig{f5}, as well as the {\em RHESSI} HXR count rates in the energy band of 12--25 keV. The {\em GOES} 1--8 \AA\ SXR flux is also plotted in \fig{f5}(a). We find that all the light curves of the flare show obvious fluctuation and exhibit certain periodicity during the rising phase (10:48--11:00 UT) of the {\em GOES} C2.8 flare (\fig{f5}(b)). The periodic variation of the flare intensity become clearer by seeing the detrended intensity fluxes (\fig{f5}(c)), which are obtained by subtracting the corresponding smoothed fluxes using a 120 s boxcar.

\begin{figure}
\epsscale{0.80}
\plotone{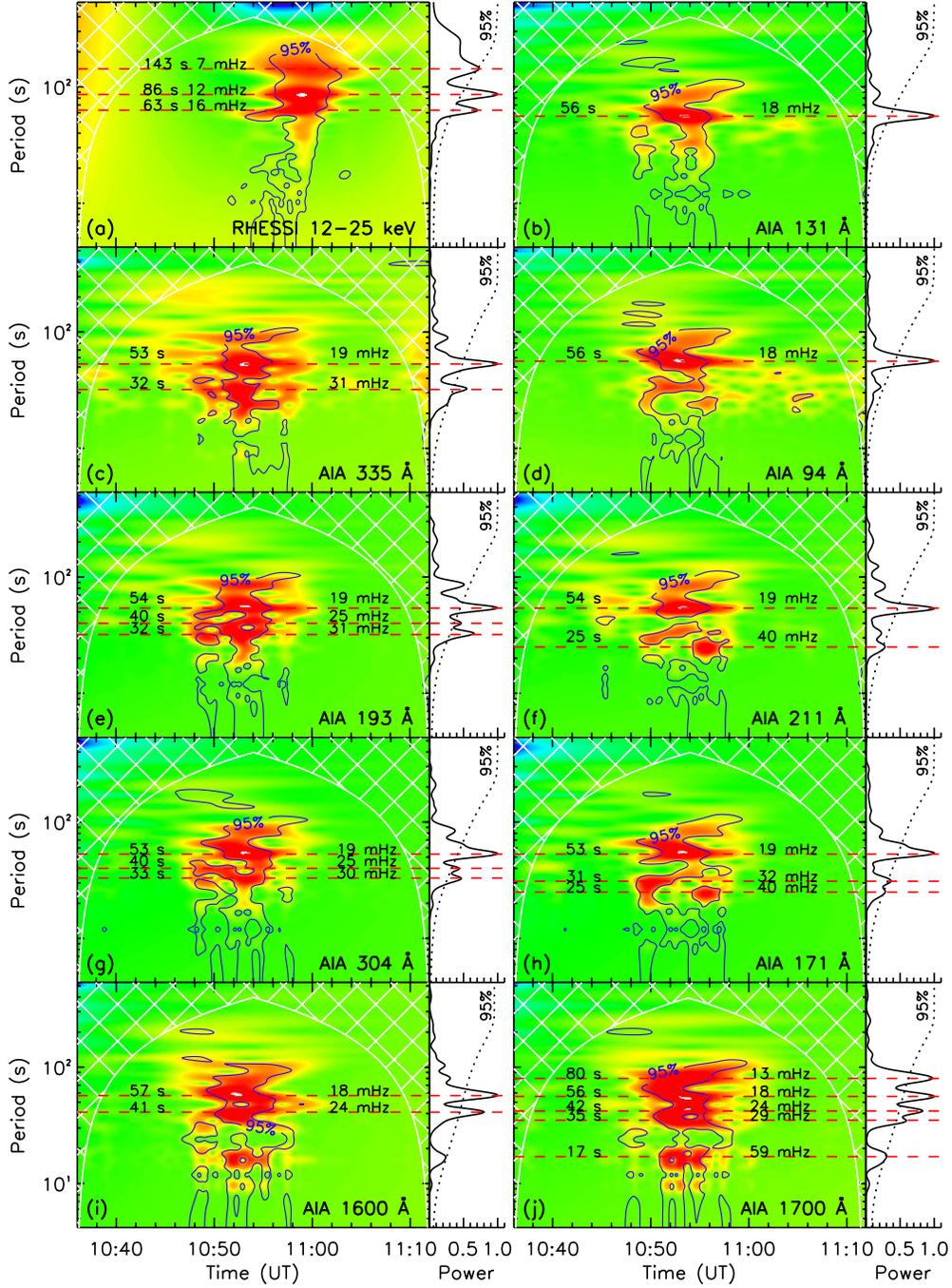}
\caption{Wavelet analysis of the detrended light curves of the flare in UV, EUV, and hard X-ray. For each wavelet power spectrum, the corresponding normalized global power is plotted on the right, in which the dotted line indicates the 95\% significance level. The red horizontal dashed lines indicate the possible periods of the flare, which are determined by the peaks (above 95\% significance level) of the global power. The blue contours in each wavelet power spectrum outline the region where the significance level is above 95\%. In this figure, redder color correspond to higher wavelet power. \label{f6}}
\end{figure}

To isolate the periods of the accompanying pulsation flare from the detrended fluxes, we resort to the wavelet analysis method, a common technique for analyzing localized variations of power within a time series. This technique allows us to investigate the time dependence periods within the observed data. The details of the procedure and the corresponding guidance are given by \cite{torr98}. In our analysis, we choose the function ``Morlet'' as the basis function, and a red-noise significance test is performed. Since both the time series and the wavelet function are finite, the wavelet can be altered by edge effects at the end of the time series. The significance of this edge effect is shown by a cone of influence (COI), defined as the region where the wavelet power drops by a factor of $e^{-2}$. Areas of the wavelet power spectrum outside the region bounded by the COI should not be included in the analysis. Our wavelet analysis results of the flare light curves are shown in \fig{f6}, in which we highlight the isolated frequencies (periods) of the QPP flare with horizontal red dashed lines. Note that the frequencies (periods) of the flare are determined by the global power where the significance is above 95\%.

Based on the wavelet analysis technique, 16 different frequencies (periods) are isolated from the AIA and {\em RHESSI} light curves of the accompanying flare, and all of them are listed in \tbl{tb}. It is interesting that a few periods (frequencies) are found simultaneously in different wavelength bands (e.g., 19 mHz (53 s)), while almost each wavelength band manifests multiple periods (frequencies) (e.g., 193 \AA). The period error of each period is determined by the full width at half maximum of each peak on the global power curve, which is obtained by fitting each peak with a Gaussian function. On the wavelet power spectrum diagrams, the blue contours represent the regions where the significance is above 95\%. We measure the duration and cycles of each period defined by these contours. The results show that the durations of these periods of the flare range from 166 s to 327 s, while the cycles range from 2.3 to 15.9. Since many measured periods are the same or showing little difference, we get the main periods by averaging the same or neighboring periods. This yields seven main periods (frequencies) of the flare (see $P_{\rm fa}$ ($F_{\rm fa}$) in \tbl{tb}).

By comparing the periods (frequencies) of the pulsation flare and the periods (frequencies) of the QFP waves, we find that five of the seven main flare periods (frequencies) coincide well with the QFP waves' periods (frequencies) which are revealed by the $k$--$\omega$ diagram shown in \fig{f4}(a). These corresponding periods (frequencies) are $25\pm2.5$ s (40 mHz), $32.6\pm3$ s (31 mHz), $41\pm3.8$ s (24.4 mHz), $55.4\pm5.8$ s (18 mHz), and $83\pm13.5$ s (12 mHz). This result indicates that the pulsation flare and the QFP waves are possibly excited by the same physical mechanism. On the other hand, the $k$--$\omega$ diagram of the QFP waves also reveals a few low frequencies (e.g., 9.3 mHz (108 s) and 2.5 mHz (400 s)) that can not be found from the pulsation flare. We conjecture that these unmatched low frequencies are possibly caused by the leakage of the photospheric $p$--mode oscillations through the chromosphere and transition region into the low corona (e.g., 5 minutes oscillation), which has been considered as an obvious source of such low frequencies in a lot of studies \citep[e.g.,][]{demo00,demo02a,mars03,depo04a,depo05,wang09a,wang09b}.

Apart from the QFP waves observed in the open loops, propagating waves along the closed transequatorial loops are also observed at both 171 \AA\ and 193 \AA\ wavelength bands. To minimize the influence from the neighbouring QFP waves that can only be observed on the AIA 171 \AA\ images, we investigate the propagating waves in the closed loops using only the AIA 193 \AA\ observations. The time-distance diagrams obtained from the AIA 193 \AA\ running difference images along the spherical curves that are shown in \fig{f1}(e) are displayed in \fig{f7}. From these time-distance diagrams, multiple steep stripes (representing the propagating wave trains) of both positive and negative slopes can be clearly observed on the time-distance diagrams (see \fig{f7}(b)). The speeds of all stripes in the plane of the sky along each cut are obtained by making a linear fit to the stripes. The average speed of all the stripes with positive slopes is \aspeed{240}, while it is \aspeed{220} for the stripes with negative slopes which represent the reflection of the propagating waves at the northern footpoints of the closed transequatorial loops. In addition, a few wave trains changed their direction at a distance of about 200 Mm from the flare kernel and their average speed is \aspeed{68} (see the thin oblique arrow in \fig{f7}(c)). The reflection of the waves suggests that they should be real MHD waves rather than the so-called ``pseudo-waves'' \citep[e.g.,][]{dela99,chen02,attr07,dela08}, which are suggested as the footprints or low coronal extensions of the associated CMEs.

\section{CONCLUSIONS AND DISCUSSIONS}
We investigate a QFP wave event in a great detail with the high temporal and spatial resolution observations taken by {\em SDO}/AIA, in which the QFP waves showed multiple arc-shaped wave trains that were propagating along a cluster of open active region coronal loops during the rising phase of the associated flare. We demonstrate that almost all the periods (frequencies) of the accompanying pulsation flare coincide well with the periods (frequencies) revealed by the $k$--$\omega$ diagram of the QFP waves. This result highly suggests that the QFP waves and the accompanying pulsation flare were possibly excited by a common physical origin. Similar result has also recently been proposed by \cite{liu11} in their report on the first QFP waves directly imaged by {\em SDO}/AIA. However, in their case, only one frequency of the QFP waves coincides with the accompanying flare. In our event, the perfect consistency of multiple frequencies of the pulsation flare and the QFP waves make us believe the existence of such a relationship between them, with a high confidence level. Furthermore, the $k$--$\omega$ diagram of the QFP waves also reveals a few low frequencies that can not be found in the flare, such as 9.3 mHz (108 s) and 2.5 mHz (400 s). We conjecture that these low frequencies are possibly the manifestations of the photospheric pressure-driven ($p$--mode) oscillations in the low corona. The leakage of the photospheric oscillations through the chromosphere and transition region into the low corona has been proved to be possible in a number of theoretical and observational studies \citep{bel77,demo00,demo02a,mars03,depo04a,depo05,chen06,wang09a,wang09b,didk11}.

It should be noted that the projection effect may affect our analysis results about the waves in the closed transequatorial loops to a certain extent. Since these loops were located within the solar limb, we can not distinguish the height of the propagating waves based on a single view point observations, since the waves might propagate along high loops or on the surface. It is possible to distinguish this problem using multi-angle observations, such as the simultaneous observations from {\em SDO} and \citep[{\em STEREO};][]{kais08}. Unfortunately, the {\em STEREO} has not the required high cadence to catch these waves. The results presented in Section 3 are based on the assumption that the waves were propagating on the surface. Here, we discuss the possibility that the waves were propagating along high loops. Assuming the angle of inclination of the high loops relative to the solar surface is $45^{\circ}$, we can obtain the propagating average speed of the waves to be \aspeed{340}, and the mean speed of the reflected waves \aspeed{310}. These speeds of the waves in the closed transequatorial loops are all above the sound speed in the low coronal. By combining the phenomena that no significant CME was observed to be associated the waves and that the waves can be reflected, we conclude that the waves should be fast magnetosonic waves rather than the so-called ``pseudo-waves'', which are thought to be the footprints or the low coronal extensions of the associated CMEs \citep[e.g.,][]{dela00,chen02,attr07}.

With the phase speeds of the propagating waves, we can calculate the local magnetic field strength of the corona structures where the waves propagated. The sound speed at formation temperature of 171 \AA\ , i.e., 0.8 MK, is about \speed{131}, which is calculated from the formula $c_{\rm s}=\sqrt{\frac{\gamma p}{\rho}}=\sqrt{\frac{2\gamma k_{\rm B} T}{\mu m_{\rm p}}}=147\sqrt{\frac{T_{\rm e}}{1 \rm MK}}$ [km s$^{-1}$] \citep{asc05}, with $\gamma=\frac{5}{3}$ being the adiabatic index, $\mu=1.27$ the mean molecular weight in corona (H:He=10:1), $m_{\rm p}=1.67 \times 10^{-24}$ g the proton mass, and $k_{\rm B}=1.38 \times 10^{-16}$ erg K$^{-1}$ the Boltzmann constant. For fast magnetosonic waves propagating along the magnetic field and the amplitude of the perturbation perpendicular to the magnetic field, the phase speed is $v_{\rm ph}=\sqrt{c_{\rm s}^2+v_{\rm A}^2}$, where $v_{\rm A}=\frac{B}{\sqrt{4 \pi \rho}}$ being the Alfv$\rm \acute{e}$n speed, $B$ the magnetic field strength, and $\rho$ the ion density. Using the phase speed $v_{\rm ph}=$ \speed{834} of the QFP waves obtained from the $k$--$\omega$ diagram of the QFP waves and the number density $n_{\rm e} \approx 10^8$ cm$^{-3}$ estimated with the 171 \AA\ channel response \citep{boer11}, we can estimate the magnetic field strength $B$ of the open active region coronal loops which guide the QFP waves, and the result is $B \approx 4.24$ Gauss. For the waves guided by the closed transequatorial loops, its phase speed $v_{\rm ph}=$ \speed{240} is about twice of the sound speed ($c_{\rm s}=$ \speed{131}). With the same method, we calculate the magnetic field strength of the closed transequatorial loops, which is $B \approx 1.24$ Gauss, about three times smaller than the magnetic field strength of the open active region coronal loops. If these waves were propagating in high loops as we discuss above, the magnetic field strength of the closed loops should be 1.6 Gauss.

Previous studies have shown that magnetic reconnection is the basic mechanism responsible for the fast release of magnetic energy in the corona. However, it is still unclear about the origin of the periodicity of solar flares and the excitation mechanism of the QFP waves. Generally, there are two alternative possibilities for the generation of the periodicity of solar flares: spontaneous, and forced periodicity \citep{naka09}. The former is determined by the local plasma properties in the reconnection site, while the latter is excited externally. For the first mechanism, magnetic energy is continuously supplied by inflow, and it continuously builds up around the vicinity of a flare epicentre. When the critical point for magnetic reconnection is reached, the stored energy is released by a burst, and the same process repeats again. A periodic regime of magnetic reconnection can be generated in several physical situations, such as the coalescence of two magnetic flux tubes \citep{taji87}, the repeated generation of plasmoids as well as their coalescence \citep{klie00}, and other mechanism \citep{ofma06,mcla09}. On the other hand, the periodicity of flares can also be produced by external agents, such as MHD waves \citep[see][and references therein]{naka09}. In the case presented here, we conjecture that the periodicity of the flare might be excited by some internal micro dynamics in the reconnection process and it thereby drives the QFP waves as well as the periodicity of the accompanying flare.

In addition, many authors found evidence that coronal waves are tightly connected with photospheric global pressure-driven ({\em p}-mode) oscillations \citep{demo00,demo02a,demo02b,mars03,mars06,chen06,wang09a,wang09b}.For {\em p}-mode oscillations in the photosphere, they are normally evanescent because their periods are well above the cutoff period in the upper photosphere and the low chromosphere as well as the transition region in nonmagnetic solar atmosphere \citep{erde07}. However, for magnetosonic gravity waves, the photospheric {\em p}-mode oscillations can leak sufficiently into the upper solar atmosphere layers, in which low frequency waves are channeled by the highly inclined magnetic fields that can increase the cutoff periods \citep{bel77}. This effect has been confirmed by the numerical simulation \citep{depo04a,depo04b,depo05} and observational studies \citep{bloo06,mcin06,vecc07,wang09a,wang09b,didk11}. In the present case, a few low frequencies are revealed by the $k$--$\omega$ diagram of the QFP waves, which do not coincide with the frequencies of the pulsation flare. Here we emphasize that the leakage of $p$--modes oscillations from the photosphere into the low corona should be a noticeable mechanism for driving QFP waves.

Following \cite{vaug05}, \cite{grub11} developed a method to test the reliability of the periodicity of several solar flares, in which both raw and detrending light curves were used to create the power spectra density (PSD) spectrum. They found that the periods revealed by the PSD spectrum made from the detrending light curve did not appear on the PSD spectrum obtained from the raw light curve, but they did reproduce the instrumental periodicity on PSD spectrum obtained from both raw and detrending light curves. They concluded that such result should be due to the under-estimating the red-noise component in PSD spectrum generated from the detrending light curves, which can introduce spurious oscillations by enhancing power of weak signals. we also use this method to test the reliability of the periodicity of the flare analyzed in this paper. We find that the PSD spectrums made from the raw light curves do show the similar periods revealed by the wavelet power spectrum, but some of them are lower than the $3\sigma$ confidence limit \citep{grub11}. In spite of this, we still believe that the periods revealed by the wavelet power spectrums, since they all lasted for several distinct cycles during the rising phase of the flare (see \tbl{tb}). If the periods of the flare are caused by the erratic, aperiodic red-noise, they should be unstable. Finally, we expect to analyze more similar events to confirm the relationship between flares and QFP waves, and theoretical work will be essential to resolve the underlying physical mechanism.

\acknowledgments {\em SDO} is a mission for NASA's Living With a Star (LWS) Program. We thank the AIA, {\em GOES}, and {\em RHESSI} teams for data support. We thank the anonymous referee for constructive comments and suggestions that improved the content of the manuscript. We also thank Dr. W. Liu, K. J. Li, and D. Gruber for their valuable suggestions and technical support on this work. The wavelet software was provided by C. Torrence and G. Compo. It is available at \url{http://atoc.colorado.edu/research/wavelets}. This work is supported by the Natural Science Foundation of China under grants 10933003, 11078004, and 11073050, and the National Key Research Science Foundation (2011CB811400).

\begin{table}
\begin{center}
\caption{All the periods, frequencies, duration, cycles, averaged periods and frequencies of the flare\label{tb}}
\begin{tabular}{llccclc}
\tableline\tableline
Wavelength &$P_{\rm f}$ (s) & $F_{\rm f}$ (mHz) &Duration &Cycles & $P_{\rm fa}$ (s) &$F_{\rm fa}$ (mHz)\\
\tableline
AIA 1700 \AA\     &$17\pm4$    &59    &264    &15.9    &$17\pm4$    &59\\
AIA 171  \AA\     &$25\pm2$    &40      &170    &6.8    &            &\\
AIA 211  \AA\     &$25\pm3$    &40    &166    &6.6    &$25\pm2.5$  &40\\
AIA 171  \AA\     &$31\pm3$    &32     &288    &7.4    &            & \\
AIA 193  \AA\     &$32\pm4$    &31     &219    &6.8    &            & \\
AIA 335  \AA\     &$32\pm2$    &31     &160    &5.0    &            & \\
AIA 304  \AA\     &$33\pm4$    &30      &246    &7.5    &            &\\
AIA 1700 \AA\     &$35\pm2$    &29    &303    &8.7    &$32.6\pm3$  &31\\
AIA 193  \AA\     &$40\pm5$    &25     &269    &6.7    &            & \\
AIA 304  \AA\     &$40\pm2$    &25    &246    &6.2    &            &  \\
AIA 1600 \AA\     &$41\pm5$    &24     &269    &6.6    &            & \\
AIA 1700 \AA\     &$42\pm3$    &24  &282    &6.7    &$41\pm3.8$  &24.4\\
AIA 171  \AA\     &$53\pm4$    &19      &296    &5.6    &            &\\
AIA 304  \AA\     &$53\pm5$    &19      &291    &5.5    &            &\\
AIA 335  \AA\     &$53\pm5$    &19     &271    &5.1    &            & \\
AIA 193  \AA\     &$54\pm9$    &19      &314    &5.8    &            &\\
AIA 211  \AA\     &$54\pm5$    &19     &305    &5.6    &            & \\
AIA 94   \AA\     &$56\pm5$    &18     &303    &5.4    &            & \\
AIA 131  \AA\     &$56\pm5$    &18     &299    &5.3    &            & \\
AIA 1700 \AA\     &$56\pm8$    &18      &303    &5.4    &            &\\
AIA 1600 \AA\     &$57\pm7$    &18     &306    &5.4    &            & \\
RHESSI 12--25 keV &$62\pm5$    &16     &307    &5.0    &            & \\
AIA 1700 \AA\     &$80\pm15$   &13    &330    &4.1    &$55.4\pm5.8$&18\\
RHESSI 12--25 keV &$86\pm12$   &12    &325    &3.8    &$83\pm13.5$ &12\\
RHESSI 12--25 keV &$143\pm19$  &7 &327    &2.3    &$143\pm19$  &7\\
\tableline
\end{tabular}
\tablecomments{In this table, $P_{\rm f}$ and $F_{\rm f}$ are the periods and frequencies of the flare in various wavelength bands, while $P_{\rm fa}$ and $F_{\rm fa}$ are the averaged periods and frequencies respectively.}
\end{center}
\end{table}

\end{document}